\documentstyle[epsf]{article}
\begin {document}
\title{Representations of the Mapping Class Group of the Two Punctured Torus on the Space of $\hat {sl}(2,C)$ Spin 1/2 - Spin 1/2 Kac-Moody Blocks}
\author{John Manolis Smyrnakis\thanks{Supported by E.C. program `Human Capital and Mobility'.}\\
Mathematical Research Institute\\
Universiteit Utrecht}
\date{November 20, 1996}
\maketitle
\begin{abstract}
The integral representations of the $\hat {sl}(2,C)$ Spin 1/2 - Spin 1/2 Kac-Moody Blocks on the torus, arising from the free field representation of the $\hat {sl}(2,C)$ Kac-Moody algebra of \cite{WA} and \cite{BF}, are used to derive an infinite class of representations of the mapping class group of the two punctured torus.  
\end{abstract}

\section{Introduction}
In recent years Wess-Zumino-Witten (WZW) models \cite{W} have received considerable attention since most known examples of conformal field theory can be undestood through them (e.g the minimal models \cite{BPZ}).  In particular the SU(2) WZW model has been examined often because it is the simplest example of non-abelian model, as well as because for level $k>1$ it provides examples of unitary conformal field theories with central charge $c>1$ (For SU(2) WZW model, c=3k/(k+2)).  It is therefore important to understand this model well at any integer level k.

The level k SU(2) WZW model is directly linked to the level k $\hat{sl}(2,C)$ Kac-Moody algebra, since the second is simply the algebra of the modes of the conserved currents of the model for each chirality.  In fact the physisists Hilbert space of this model, is a direct sum of a finite number of irreducible $\hat{sl}(2,C)$ Kac-Moody modules for each chiral sector, indexed by their spin which takes values from 0 to k/2, as specified by the selection rules. Here we are going to denote these modules by $H_j$.

The SU(2) WZW model, like any other WZW model, can be formulated on a Riemann 
surface of arbitrary genus.  The simplest case of the sphere is well understood \cite{KZ},\cite{DF}.  Here we are going to be concerned with the case of the torus.  Of particular interest to us is going to be the space of the two-point 
spin1/2-spin1/2 blocks on the torus. This space is generated by the following basis elements: 
$$F^{+}_{2j+1}(z,\tau )\equiv const\cdot Tr_{H_j}(q^{L_0-c/24}\cdot  _{-1/2}\Phi^{1/2}_{j,j+1/2}(z)
\cdot _{1/2}\Phi^{1/2}_{j+1/2,j}(0))$$
$$F^{-}_{2j+1}(z,\tau )\equiv const\cdot Tr_{H_j}(q^{L_0-c/24}\cdot  _{-1/2}\Phi^{1/2}_{j,j-1/2}(z)
\cdot _{1/2}\Phi^{1/2}_{j-1/2,j}(0))$$
Here $$_\mu\Phi^{1/2}_{j_1,j_2}:\\  H_{j_2}\rightarrow H_{j_1}$$
are the chiral primary fields \cite{BF}, $L_0$ is the diagonal Virasoro algebra 
generator, in our case determined from the current modes through the 
Sugawara construction, and $q=e^{2\pi i\tau }$.  z takes values on the parallelogram with sides 1 and $\tau $, $Im\tau >0$.  

These basis elements for the above space of blocks can be calculated via the use of 
a free field representation \cite{WA},\cite{BF}.  For our basis elements this calculation has been carried out explicitely in \cite{S}.  $F_n^+(z,\tau)$ is given by the following integral  
$$\begin{array}{ll} F_n^+(z,\tau) &={const\over \eta^3(\tau)}\cdot \int_0^1dt
\Theta  \left( \begin{array}{c}{n\over 2p}\\ 0 \end{array} \right) (-z-2t,2p\tau )
E(t,\tau )^{-1\over p} E(z+t,\tau)^{-1\over p}E(-z,\tau)^{1\over 2p} \\ 
 & \cdot {d^2\over dt^2}ln{E(z+t,\tau)\over E(t,\tau)} \end{array} $$
while $F_n^-(z,\tau )=F_{-n}^+(z,\tau )$.  Here 
$$\Theta  \left( \begin{array}{c}{n\over 2p}\\ 0 \end{array} \right) (z,2p\tau )=
\sum_{l=-\infty }^\infty e^{\pi i(l+{n\over 2p})^22p\tau }e^{2\pi i(l+{n\over 2p})z}$$
 is a theta function with characteristics \cite{F}, and
$$E(z,\tau )={\Theta_1(z,\tau )\over \Theta_1'(0,\tau )}$$
is the prime form on the torus.  Here $\Theta_1(z,\tau )$ is the first Jacobi theta function and $\eta (\tau )$ is the Jacobi eta function.  

Our vector space of blocks forms the fiber of a flat holomorphic vector bundle over the punctured torus with respect to the Knizhnik-Zamolodchikov connection.  In fact our basis is a horizondal section of this vector bundle.  So there are monodromy matrices associated to analytic continuation of our basis elements around non-contractible cycles a and b, and we will denote them by a,b respectively (here the constant in front of the blocks is taken to be 1). There is also a matrix $\sigma $ associated to analytic continuation from z to -z.  Since the blocks are translation invariant, analytically continuing from z to -z is the same as interchanging the origin with z, so this is really a braiding matrix.  This braiding matrix was originally obtained in \cite{TK} through a calculation 
on the sphere.  Together with a,b they form a translation invariant representation of the braid group of two strings on the torus \cite{S}.  But the space of conformal blocks only depends on the complex 
structure of the torus so the blocks transform covariantly under appropriate lifts of the SL(2,Z) transformations on the unpunctured torus, on the mapping class group of the two punctured torus.  The action of these lifts, denoted S and T respectively, on the space of our conformal blocks, is determined here.  In this way an infinite set of representations of the mapping class group of the two punctured torus is obtained.    
 
The organization of this paper is as follows:  First the structure of the 1 forms that are used to obtain the blocks is examined and the identities satisfied by the various integrals of these forms are written down.  The selection rules for the two point blocks  follow from these identities.  Then these identities are used to obtain the S matrix, corresponding to $(z,\tau )\rightarrow (z/\tau ,-1/\tau )$.  After that a presentation of the mapping class group is obtained by making use of its action on the fundamental group of the two 
punctured torus as outer automorphisms. Then a set of identities satisfied by the computed matrices is obtained through analytic continuation.  These identities are checked for level 1,2,3 
and they are satisfied by the computed matrices.  Finally it is shown that the computed matrices form a particular class of representations of the mapping class group of the two punctured torus.  

\section{Structure of the differential forms giving the conformal blocks}
The forms of interest here are the following:
$$\begin{array}{ll} \omega_\mu (t,z,\tau) &={1\over \eta^3(\tau)}\cdot 
\Theta  \left( \begin{array}{c}{\mu \over 2p}\\ 0 \end{array} \right) (-z-2t,2p\tau )
E(t,\tau )^{-1\over p} E(z+t,\tau)^{-1\over p}E(-z,\tau)^{1\over 2p} \\ 
 & \cdot {d^2\over dt^2}ln{E(z+t,\tau)\over E(t,\tau)}dt \end{array} $$
This is defined on an appropriately cut universal cover of the torus.  Here the 
cuts are chosen as follows:

\begin{picture}(105,100)
\setlength{\unitlength }{0.02in}

\put (0,0){
\begin{picture}(35,30)
\put (0,0){\line(1,4){5}}
\put (0,0){\line(1,0){30}}
\put (30,0){\line(1,4){5}}
\put (5,20){\line(1,0){30}}
\put (10,10){\circle*{2}}
\multiput(10,10)(-2,-2){5}{\line(0,-1){1}}
\multiput(8,8)(-2,-2){5}{\line(0,1){1}}
\multiput(8,9)(-2,-2){5}{\line(1,0){2}}
\multiput(5,20)(1,-2){5}{\line(0,-1){2}}
\multiput(5,18)(1,-2){5}{\line(1,0){1}}
\put(8,12){-z-$\tau $-1}
\end{picture}
}

\put (30,0){
\begin{picture}(35,30)
\put (0,0){\line(1,4){5}}
\put (0,0){\line(1,0){30}}
\put (30,0){\line(1,4){5}}
\put (5,20){\line(1,0){30}}
\put (10,10){\circle*{2}}
\multiput(10,10)(-2,-2){5}{\line(0,-1){1}}
\multiput(8,8)(-2,-2){5}{\line(0,1){1}}
\multiput(8,9)(-2,-2){5}{\line(1,0){2}}
\multiput(5,20)(1,-2){5}{\line(0,-1){2}}
\multiput(5,18)(1,-2){5}{\line(1,0){1}}
\put(8,12){-z-$\tau $}
\end{picture}
}

\put (60,0){
\begin{picture}(35,30)
\put (0,0){\line(1,4){5}}
\put (0,0){\line(1,0){30}}
\put (30,0){\line(1,4){5}}
\put (5,20){\line(1,0){30}}
\put (10,10){\circle*{2}}
\multiput(10,10)(-2,-2){5}{\line(0,-1){1}}
\multiput(8,8)(-2,-2){5}{\line(0,1){1}}
\multiput(8,9)(-2,-2){5}{\line(1,0){2}}
\put(8,12){-z-$\tau $+1}
\end{picture}
}

\put (5,20){
\begin{picture}(35,30)
\put (0,0){\line(1,4){5}}
\put (0,0){\line(1,0){30}}
\put (30,0){\line(1,4){5}}
\put (5,20){\line(1,0){30}}
\put (10,10){\circle*{2}}
\multiput(10,10)(-2,-2){5}{\line(0,-1){1}}
\multiput(8,8)(-2,-2){5}{\line(0,1){1}}
\multiput(8,9)(-2,-2){5}{\line(1,0){2}}
\multiput(5,20)(1,-2){5}{\line(0,-1){2}}
\multiput(5,18)(1,-2){5}{\line(1,0){1}}
\put(8,12){-z-1}
\end{picture}
}

\put (35,20){
\begin{picture}(35,30)
\put (0,0){\line(1,4){5}}
\put (0,0){\line(1,0){30}}
\put (30,0){\line(1,4){5}}
\put (5,20){\line(1,0){30}}
\put (10,10){\circle*{2}}
\multiput(10,10)(-2,-2){5}{\line(0,-1){1}}
\multiput(8,8)(-2,-2){5}{\line(0,1){1}}
\multiput(8,9)(-2,-2){5}{\line(1,0){2}}
\put(8,12){-z}
\end{picture}
}

\put (65,20){
\begin{picture}(35,30)
\put (0,0){\line(1,4){5}}
\put (0,0){\line(1,0){30}}
\put (30,0){\line(1,4){5}}
\put (5,20){\line(1,0){30}}
\put (10,10){\circle*{2}}
\multiput(10,10)(-2,-2){5}{\line(0,-1){1}}
\multiput(8,8)(-2,-2){5}{\line(0,1){1}}
\multiput(8,9)(-2,-2){5}{\line(1,0){2}}
\multiput(5,20)(1,-2){5}{\line(0,-1){2}}
\multiput(5,18)(1,-2){5}{\line(1,0){1}}
\put(8,12){-z+1}
\end{picture}
}

\put (10,40){
\begin{picture}(35,30)
\put (0,0){\line(1,4){5}}
\put (0,0){\line(1,0){30}}
\put (30,0){\line(1,4){5}}
\put (5,20){\line(1,0){30}}
\put (10,10){\circle*{2}}
\multiput(10,10)(-2,-2){5}{\line(0,-1){1}}
\multiput(8,8)(-2,-2){5}{\line(0,1){1}}
\multiput(8,9)(-2,-2){5}{\line(1,0){2}}
\put(8,12){-z+$\tau $-1}
\end{picture}
}

\put (40,40){
\begin{picture}(35,30)
\put (0,0){\line(1,4){5}}
\put (0,0){\line(1,0){30}}
\put (30,0){\line(1,4){5}}
\put (5,20){\line(1,0){30}}
\put (10,10){\circle*{2}}
\multiput(10,10)(-2,-2){5}{\line(0,-1){1}}
\multiput(8,8)(-2,-2){5}{\line(0,1){1}}
\multiput(8,9)(-2,-2){5}{\line(1,0){2}}
\multiput(5,20)(1,-2){5}{\line(0,-1){2}}
\multiput(5,18)(1,-2){5}{\line(1,0){1}}
\put(8,12){-z+$\tau $}
\end{picture}
}

\put (70,40){
\begin{picture}(35,30)
\put (0,0){\line(1,4){5}}
\put (0,0){\line(1,0){30}}
\put (30,0){\line(1,4){5}}
\put (5,20){\line(1,0){30}}
\put (10,10){\circle*{2}}
\multiput(10,10)(-2,-2){5}{\line(0,-1){1}}
\multiput(8,8)(-2,-2){5}{\line(0,1){1}}
\multiput(8,9)(-2,-2){5}{\line(1,0){2}}
\multiput(5,20)(1,-2){5}{\line(0,-1){2}}
\multiput(5,18)(1,-2){5}{\line(1,0){1}}
\put(8,12){-z+$\tau $+1}
\end{picture}
}

\end{picture}

\noindent Here each line of cuts consists of $p-1$ segments
 
This cut plane has two automorphisms, $s=\tau -z-t$ and $s=1-z-t$.  This means that if we do this changes of variable then we do not separate any regions on the cut plane.  Hence we can derive the following identities:
\newtheorem{lemma}{Lemma}
\begin{lemma}
The following identities hold: 
\begin{equation}
\int_{-z}^{-z+1}\omega_\mu (t,z,\tau )dt=-\tilde q^{-\mu}\int_0^1\omega_{-\mu}(t,z,\tau)dt
\end{equation}
\begin{equation}
\int_{-z}^{-z+\tau}\omega_\mu (t,z,\tau )dt=-\int_0^\tau\omega_{2-\mu}(t,z,\tau)dt
\end{equation}
\begin{equation}
\int_1^{-z+1}\omega_\mu (t,z,\tau )dt=\tilde q^{-\mu}\int_0^{-z}\omega_{\mu}(t,z,\tau)dt
\end{equation}
\begin{equation}
\int_{\tau}^{-z+\tau}\omega_\mu (t,z,\tau )dt=\int_0^{-z}\omega_{\mu -2}(t,z,\tau)dt
\end{equation}
where $\tilde q=e^{2\pi i\over p}$.  
\end{lemma}
{\bf Proof:} For the first identity, let s=1-z-t. Then 
$$\int_{-z}^{-z+1}dt \cdot \omega_\mu (t,z,\tau )=\int_0^1ds \cdot \omega_\mu (1-z-s,z,\tau)$$
But $$\Theta  \left( \begin{array}{c}{\mu \over 2p}\\ 0 \end{array} \right) (z+2s-2,2p\tau )=
\tilde q^{-\mu}\Theta  \left( \begin{array}{c}{-\mu \over 2p}\\ 0 \end{array} \right) (-z-2s,2p\tau ),$$
$$E(1-z-s)^{-1\over p}=E(z+s)^{-1\over p}$$
$$E(1-s)^{-1\over p}=E(s)^{-1\over p}$$
where for the last two analytic continuations the path is a straight line.  
Putting everything together we get that  
$\omega_\mu (1-z-s,z,\tau)=-\tilde q^{-\mu} \omega_{-\mu} (s,z,\tau)$, so the 
first identity is proven.  To prove the second identity one needs the change of 
variable $s=\tau -z-t$.  For the third and forth it is only necessary to analytically 
continue.  

The above identities are the identities of the integrals that we can derive directly from 
the transformation properties of the theta function and the prime form.  But it is possible 
to derive more independent identities by simply considering contractible loops in the 
cut plane.  Here we will consider the following loops: 

\begin{picture}(200,50)
\put(0,0){
\begin{picture}(45,20)
\setlength{\unitlength }{0.02in}
\put(3,1){\line(1,1){9}}
\put(3,1){\line(1,0){30}}
\put(33,1){\line(1,1){9}}
\put(12,10){\line(1,0){26}}
\put(40,10){\oval(4,4)[b]}

\put (0,0){\line(1,4){5}}
\put (0,0){\line(1,0){30}}
\put (30,0){\line(1,4){5}}
\put (5,20){\line(1,0){30}}
\put (10,10){\circle*{2}}
\put (40,10){\circle*{2}}
\multiput(10,10)(-2,-2){5}{\line(0,-1){1}}
\multiput(8,8)(-2,-2){5}{\line(0,1){1}}
\multiput(8,9)(-2,-2){5}{\line(1,0){2}}
\put(8,12){-z}
\put(38,12){-z+1}
\end{picture}
}

\put(80,0){
\begin{picture}(35,35)
\setlength{\unitlength }{0.02in}
\put(2,4){\line(1,4){5}}
\put(2,4){\line(1,1){8}}
\put(10,12){\line(1,4){4}}
\put(7,24){\line(1,1){8}}
\put(15,30){\oval(4,4)[l]}

\put (0,0){\line(1,4){5}}
\put (0,0){\line(1,0){30}}
\put (30,0){\line(1,4){5}}
\put (5,20){\line(1,0){30}}
\put (10,10){\circle*{2}}
\put (15,30){\circle*{2}}
\multiput(10,10)(-2,-2){5}{\line(0,-1){1}}
\multiput(8,8)(-2,-2){5}{\line(0,1){1}}
\multiput(8,9)(-2,-2){5}{\line(1,0){2}}
\put(8,12){-z}
\put(17,30){-z+$\tau $}
\end{picture}
}

\put(160,0){
\begin{picture}(35,35)
\setlength{\unitlength }{0.02in}
\put(2,4){\line(1,4){3.5}}
\put(2,4){\line(1,1){8}}
\put(5.5,18){\line(1,0){28}}
\put(11,11){\circle*{1}}
\put(10.5,11.5){\circle*{1}}
\put(11.5,10.5){\circle*{1}}
\put(10,12){\circle*{1}}
\put(12,10){\circle*{1}}
\put(4,2){\line(1,1){8}}
\put(4,2){\line(1,0){25.5}}
\put(29.5,2){\line(1,4){4}}

\put (0,0){\line(1,4){5}}
\put (0,0){\line(1,0){30}}
\put (30,0){\line(1,4){5}}
\put (5,20){\line(1,0){30}}
\put (10,10){\circle*{2}}
\multiput(10,10)(-2,-2){5}{\line(0,-1){1}}
\multiput(8,8)(-2,-2){5}{\line(0,1){1}}
\multiput(8,9)(-2,-2){5}{\line(1,0){2}}
\put(15,10){-z}
\end{picture}
}

\end{picture}

\noindent  Demanding that the integral of our form $\omega_\mu (t,z,\tau)$ around these 
loops be zero and using the identities of lemma 1 we get the following new identities: 
\begin{lemma}
The following identities hold:
\begin{equation}
\int_0^1dt\cdot \omega_\mu (t,z,\tau)+\tilde q^{-\mu }\int_0^1dt\cdot \omega_{-\mu} (t,z,\tau)
=(1-\tilde q^{-\mu})\int_0^{-z}dt\cdot \omega_\mu (t,z,\tau)
\end{equation}
\begin{equation}
\int_0^{\tau }dt\cdot \omega_\mu (t,z,\tau)+\int_0^{\tau }dt\cdot \omega_{2-\mu } (t,z,\tau)=
\tilde q^{-1} \left( \int_0^{-z }dt\cdot \omega_\mu (t,z,\tau)-
\int_0^{-z }dt\cdot \omega_{\mu -2}(t,z,\tau)\right)
\end{equation}
\begin{equation}
\begin{array} {c}
(1-\tilde q^{-\mu +1})\int_0^{\tau }dt\cdot \omega_\mu (t,z,\tau )=
\int_0^1dt\cdot \omega_\mu (t,z,\tau )-
\tilde q^{-1}\int_0^1dt\cdot \omega_{\mu -2}(t,z,\tau )+ \\ 
(\tilde q^{-1}-1)\int_0^{-z}dt\cdot \omega_\mu (t,z,\tau )
\end{array}
\end{equation}
where $\int_0^{-z}$ is the integral below the branch cut.  Note that identities (5) 
and (7) imply (6) only if $\mu \neq 1 mod p$.
\end{lemma}

The factor $\tilde q^{-1}$ in identities (6) and (7) appear because the integral from 0 to -z has to be moved below the branch cut.  These identities show that we can express all integrals 
of our forms in terms of the integrals from 0 to 1.  In fact solving for the integrals from 
0 to $\tau $ we get the following identity:
\begin{equation}
\begin{array}{c}\int_0^{\tau }dt\cdot \omega_\mu (t,z,\tau)= 
-{\tilde q^{\mu -1}\over 1-\tilde q^{\mu }}\int_0^1dt\cdot \omega_\mu (t,z,\tau)+
{\tilde q^{\mu -2}\over 1-\tilde q^{\mu -1}}\int_0^1dt\cdot \omega_{\mu -2}(t,z,\tau)+\\ 
{\tilde q^{\mu -2}(1-\tilde q)\over (1-\tilde q^{\mu -1})(1-\tilde q^{\mu })}\int_0^1dt\cdot \omega_{-\mu }(t,z,\tau) \end{array}
\end{equation}
where $\mu $ cannot take the values 0 and 1 mod p.

The identities that we have imply also that $\int_0^1dt\cdot \omega_\mu (t,z,\tau)=0$
 for $\mu =0$ mod p.  To see this take  $\mu =0$ in (5).  This way we get: 
$$\int_0^{-z}dt\cdot \omega_{0}(t,z,\tau )=-\int_0^{-z}dt\cdot \omega_{0}(t,z,\tau )$$
Also equations (5) and (7) imply that $\int_0^1dt\cdot \omega_\mu (t,z,\tau)=0$ for 
$\mu =-1$ mod p.  The vanishing of these blocks is of course to be expected from the 
representation point of view because the intermediate state in our trace that defines the 
blocks cannot have spin smaller than 0 or greater than k/2.  

\section{The modular transformation matrices}
The conformal blocks that we have computed carry a representation of a lift of the mapping class group 
of the nonpunctured torus, that is SL(2,Z), on the mapping class group of the two punctured torus.  The action of this lift on the space of conformal 
blocks is generated by two elements.  One that sends $(z,\tau )\rightarrow (z,\tau +1)$,
the T element,  
and one that sends $(z,\tau )\rightarrow (z/\tau ,-1/\tau)$, the S element.  For the second one one has 
to specify also the path of analytic continuation in the z-plane. This is going to be 
winding clockwise around the origin.  

The T matrix is easy to compute.  Letting $\tau \rightarrow \tau+1$ in the trace formuli given for the blocks in the introduction, we get a phase factor appearing in front of 
each block.  In particular $F_{2j+1}^+(z,\tau +1)=\tilde q^{j(j+1)-k/8}F_{2j+1}^+(z,\tau )$
which, using p=k+2 and n=2j+1 instead of k and j becomes:
$$F_n^+(z,\tau +1)=\tilde q^{n^2/4-p/8}F_n^+(z,\tau )$$
Using also $F_n^-(z,\tau )=F_{-n}^+(z,\tau )$ we get the same factor in front of $F_n^-(z,\tau )$.  
So if we choose as our basis 
\begin{equation} 
[F_{p-1}^-(z,\tau ),F_{p-2}^+(z,\tau ),F_{p-2}^-(z,\tau ),\ldots ,
F_2^+(z,\tau ),F_2^-(z,\tau ),F_1^+(z,\tau )]
\end{equation}
the T matrix becomes:
\begin{equation}
T=diag\{ \tilde q^{(p-1)^2/4-p/8},\tilde q^{(p-2)^2/4-p/8},\tilde q^{(p-2)^2/4-p/8},\ldots 
\tilde q^{2^2/4-p/8},\tilde q^{2^2/4-p/8},\tilde q^{1^2/4-p/8}\} 
\end{equation}

Computing the S matrix is not so easy.  Here we need the integral representation of the 
blocks.  Letting the transformation defining the S matrix act on the block $F_n^+(Z,\tau )$ we have:
$$\begin{array}{ll} F_n^+({z\over \tau },{-1\over \tau}) &={1\over \sqrt{-i\tau}^3\eta^3(\tau)}\cdot \int_0^1dt
\Theta  \left( \begin{array}{c}{n\over 2p}\\ 0 \end{array} \right)
 (-{z\over \tau}-2t,-{2p\over \tau })
E(t,{-1\over \tau })^{-1\over p} E({z\over \tau }+t,{-1\over \tau })^{-1\over p} \\
& \cdot E(-{z\over \tau},{-1\over \tau })^{1\over 2p}  
 {d^2\over dt^2}ln{E({z\over \tau}+t,{-1\over \tau})\over E(t,{-1\over \tau})} \end{array} $$
Here we have made use of the transformation property $\eta ({-1\over \tau })=
\sqrt{-i\tau }\eta (\tau )$. 
The integration contour is:
\begin{picture}(35,50)
\put(0,20){
\begin{picture}(35,30)
\setlength{\unitlength }{0.02in}
\put (0,0){\line(1,4){5}}
\put (0,0){\line(1,0){30}}
\put (30,0){\line(1,4){5}}
\put (5,20){\line(1,0){30}}
\put (5,-5){\circle*{2}}
\put(-5,-11){$-z\over \tau $}
\put(15,0){\oval(30,14)[b]}
\end{picture}
}
\end{picture}

Doing now the transformation $t=s/\tau $ we get: 
$$\begin{array}{ll} F_n^+({z\over \tau },{-1\over \tau}) &={1\over \sqrt{-i\tau}^3\eta^3(\tau)}\cdot \int_0^{\tau }{ds\over \tau }
\Theta  \left( \begin{array}{c}{n\over 2p}\\ 0 \end{array} \right)
 (-{z+2s\over \tau},-{2p\over \tau })
E({s\over \tau },{-1\over \tau })^{-1\over p} E({z+s\over \tau },{-1\over \tau })^{-1\over p} \\
& \cdot E(-{z\over \tau},{-1\over \tau })^{1\over 2p}  
 \tau ^2{d^2\over ds^2}ln{E({z+s\over \tau},{-1\over \tau})\over E({s\over \tau },{-1\over \tau})} \end{array} $$
\noindent Here the integration contour is: 
\begin{picture}(30,30)
\setlength{\unitlength }{0.002in}
\put (0,0){\line(1,4){50}}
\put (0,0){\line(1,0){300}}
\put (300,0){\line(1,4){50}}
\put (50,200){\line(1,0){300}}
\put (0,0) { \line (1,1) {100} }
\put (120,100){\line(-1,2){50}}
\multiput(100,100)(-20,-20){5}{\line(0,-1){10}}
\multiput(80,80)(-20,-20){5}{\line(0,1){10}}
\multiput(80,90)(-20,-20){5}{\line(1,0){20}}
\put (120,100){-z}
\put (100,100){\circle*{5}}
\end{picture}

The theta function and the prime form satisfy the following transformation identities, 
which are based on Poisson resummation:   
\begin{equation}
\Theta  \left( \begin{array}{c}{n\over 2p}\\ 0 \end{array} \right)
 ({z\over \tau},-{2p\over \tau })=
\sqrt{-i\tau \over 2p}e^{\pi i{z^2\over 2p\tau }}
\sum_{\lambda =0}^{2p-1}e^{2\pi in\lambda \over 2p}
\Theta  \left( \begin{array}{c}{-\lambda \over 2p}\\ 0 \end{array} \right)
 (z,2p\tau ) 
\end{equation}
\begin{equation}
E({z\over \tau },{-1\over \tau })^\alpha =
{e^{\pi i\alpha z^2\over \tau }\over \tau ^\alpha }E(z,\tau )^\alpha 
\end{equation}

Using now these identities together with the periodicity of the theta functions 
in $\lambda $ with period 2p, we get the following expression:  
\begin{equation}
F_n^+({z\over \tau },{-1\over \tau })=
{i\tau ^{3\over 2p}\over \sqrt{2p}}
\sum_{\mu =1}^{2p}\tilde q^{-n\mu \over 2}\int_{C0}^\tau ds\omega_\mu (s,z,\tau )
\end{equation}
where C is the integration contour in the previous picture.  

Now  we can deform this contour to one that goes from 0 around -z back to 0 on the 
other side of the branch cut and then from there to $\tau $.  Doing this we get:  
$$\int_{C0}^\tau ds\omega_\mu (s,z,\tau )=\int_0^\tau ds\omega_\mu (s,z,\tau )+
(1-\tilde q^{-1})\int_0^{-z}ds\omega_\mu (s,z,\tau )$$
Here the integral from 0 to -z is the one below the branch cut.  
Putting this back in equation (14) we get:
\begin{equation}
F_n^+({z\over \tau },{-1\over \tau })=\Sigma_1(z,\tau )+\Sigma_2(z,\tau )
\end{equation}
where
$$\Sigma_1(z,\tau )=
{i\tau ^{3\over 2p}(1-\tilde q^{-1})\over \sqrt{2p}}
\sum_{\mu =1}^{2p}\tilde q^{-n\mu \over 2}\int_{0}^{-z} ds\omega_\mu (s,z,\tau )$$
$$\Sigma_2(z,\tau )=
{i\tau ^{3\over 2p}\over \sqrt{2p}}
\sum_{\mu =1}^{2p}\tilde q^{-n\mu \over 2}\int_0^\tau ds\omega_\mu (s,z,\tau )$$

Substituting in the expression for $\Sigma_1(z,\tau )$ from identity (5) and in $\Sigma_2(z,\tau )$ from (8) we get:  
$$\Sigma_1(z,\tau )=
{i\tau ^{3\over 2p}\over \sqrt{2p}}\tilde q^{-1}(1-\tilde q)
\sum_{\mu =1}^{p-2}\tilde q^{-n\mu \over 2}\tilde q^\mu 
{1-\tilde q^{n\mu }\over 1-\tilde q^\mu }(F_\mu ^+(z,\tau )+(-1)^nF_{p-\mu}^-(z,\tau ))$$
$$\begin {array}{ll} 
\Sigma_2(z,\tau )&=
{i\tau ^{3\over 2p}\over \sqrt{2p}}
\sum_{\mu =1}^{p-2}\tilde q^{-n\mu \over 2}
{-\tilde q^{\mu-1}+\tilde q^{2\mu }+\tilde q^{\mu -n}-\tilde q^{2\mu -n}+
\tilde q^{n\mu +\mu -1}-\tilde q^{n\mu +\mu }\over 
(1-\tilde q^\mu )(1-\tilde q^{\mu +1})}\cdot \\ 
&(F_\mu ^+(z,\tau )+(-1)^nF_{p-\mu}^-(z,\tau ))
\end {array} $$
Putting these together we get:
\begin{equation}
\begin {array}{ll}
F_n^+({z\over \tau },{-1\over \tau })=
&{i\tau ^{3\over 2p}\over \sqrt{2p}}
\sum_{\mu =1}^{p-2}\tilde q^{-n\mu \over 2}
{\tilde q^{\mu -n}-\tilde q^{2\mu -n}+\tilde q^{n\mu +2\mu }-\tilde q^{\mu }+
\tilde q^{2\mu +1}-\tilde q^{n\mu +2\mu +1}\over 
(1-\tilde q^\mu )(1-\tilde q^{\mu +1})}\cdot  \\ 
&(F_\mu ^+(z,\tau )+(-1)^nF_{p-\mu}^-(z,\tau ))
\end {array} 
\end {equation}
The factor $ \tau ^{3/2p} $ cancells because the blocks are really densities 
in z of weight 3/2p and we have done the transformation $z\rightarrow z/\tau $.  

Now to describe the S matrix we need to introduce the matrix elements 
\begin{equation}
C_{n\mu }={i\over \sqrt{2p}}
\tilde q^{-n\mu \over 2}
{\tilde q^{\mu -n}-\tilde q^{2\mu -n}+\tilde q^{n\mu +2\mu }-\tilde q^{\mu }+
\tilde q^{2\mu +1}-\tilde q^{n\mu +2\mu +1}\over 
(1-\tilde q^\mu )(1-\tilde q^{\mu +1})}
\end{equation}
and the matrix K that has everywhere 0's except at the positions (1,k+1), (2,k), 
(3,k+2), (4,k-1), (5,k+3), (6,k-2) up to (2k-1,2k), (2k,1), where it has ones.  
This matrix is the transformation matrix from the basis 
\begin{equation} 
[F_1^+(z,\tau ), F_2^+(z,\tau ), \ldots ,F_{p-2}^+(z,\tau ), F_{p-1}^-(z,\tau), F_{p-2}^-(z,\tau), \ldots F_2^-(z,\tau )]
\end{equation}
to our original basis (9).  
In terms of the basis (17) the S matrix becomes:
$$S\prime =
\left(
\begin{array}{cc}
(C_{n\mu }) & ((-1)^nC_{n\mu })\\ 
((-1)^\mu C_{n\mu }) & ((-1)^{p+\mu +n}C_{n\mu })
\end{array}
\right) $$
where n and $\mu $ range from 1 to p-2, so S is a 2p-4 by 2p-4  matrix.  
So the S matrix in terms of the basis (9) is 
\begin{equation}
S=KS\prime K^T.
\end{equation} 

Also in terms of this basis the matrices $a,b,\sigma $ corresponding to  
the transformations 
$z+1\rightarrow z$, $z\rightarrow z-\tau $, $-z\rightarrow z$, where the analytic 
continuation for the first two transformations is along straight paths and in the 
third along a path that winds clockwise around the origin, become \cite{S}:  
$$a=\tilde q^{1\over 4}diag(\tilde q^{k\over 2}, \tilde q^{-{k-1\over 2}-1}, 
\tilde q^{k-1\over 2},  \tilde q^{-{k-2\over 2}-1}, \tilde q^{k-2\over 2}, 
\ldots ,\tilde q^{-{1\over 2}-1}, \tilde q^{1\over 2}, \tilde q^{-1})$$

$$b=\tilde q^{-{1\over 4}}\left( 
\begin{array}{ccccccccc}

0&\tilde q^{-{1\over 2}}&0&0&0&0&\ldots&0&0\\
 b_{11}(p-2)&0&0&b_{12}(p-2)&0&0&\ldots &0&0\\
b_{21}(p-2)&0&0&b_{22}(p-2)&0&0&\ldots &0&0\\
0&0&b_{11}(p-3)&0&0&b_{12}(p-3)&\ldots &0&0\\
0&0&b_{21}(p-3)&0&0&b_{22}(p-3)&\ldots &0&0\\
\vdots &\vdots &\vdots &\vdots &\vdots &\vdots &\ddots &\vdots &\vdots \\
0&0&0&0&0&0&\ldots &0&b_{12}(2)\\
0&0&0&0&0&0&\ldots &0&b_{22}(2)\\
0&0&0&0&0&0&\ldots &\tilde q^{-{1\over 2}}&0

\end{array}
\right)$$

$$\sigma =\tilde q^{-{3\over 4}}
\left(
\begin{array}{ccccccc}
1&0&0&0&0&\ldots &0\\
0&b(k)&c(k)&0&0&\ldots &0\\
0&c(-k)&b(-k)&0&0&\ldots &0\\
0&0&0&b(k-1)&c(k-1)&\ldots &0\\
0&0&0&c(-k+1)&b(-k+1)&\ldots &0\\
\vdots &\vdots &\vdots &\vdots &\vdots &\ddots &\vdots \\
0&0&0&0&0&\ldots &1
\end{array}
\right)$$
where $b_{11}(n)= -\tilde q^{-n-{1\over 2}}b(-n)$, 
$b_{12}(n)=\tilde q^{n-{1\over 2}}c(n)$, 
$b_{21}(n)=\tilde q^{-n-{1\over 2}}c(-n)$, $b_{22}(n)= -\tilde q^{n-{1\over 2}}b(n)$ 
and $b(n)={\tilde q-1\over \tilde q^n-1}$, 
$c(n)={\tilde q^{-n}(\tilde q^{n+1}-1)\over \tilde q^n-1}$.

\section{A presentation of the mapping class group of the two punctured torus}
Let us denote this mapping class group by ${\cal M}_{1,2}$.  Here 1 stands for the genus and 2 for the number of punctures.  Similarly let $B_{1,2}$ denote the full braid group of two strings on the torus.  Then we have the following short exact sequence \cite{B}:  $$1\rightarrow B_{1,2}/center\rightarrow {\cal M}_{1,2}\rightarrow SL(2,Z)\rightarrow 1$$
$B_{1,2}/center$ is generated by the following generators:

\epsfxsize =\textwidth 
\epsffile{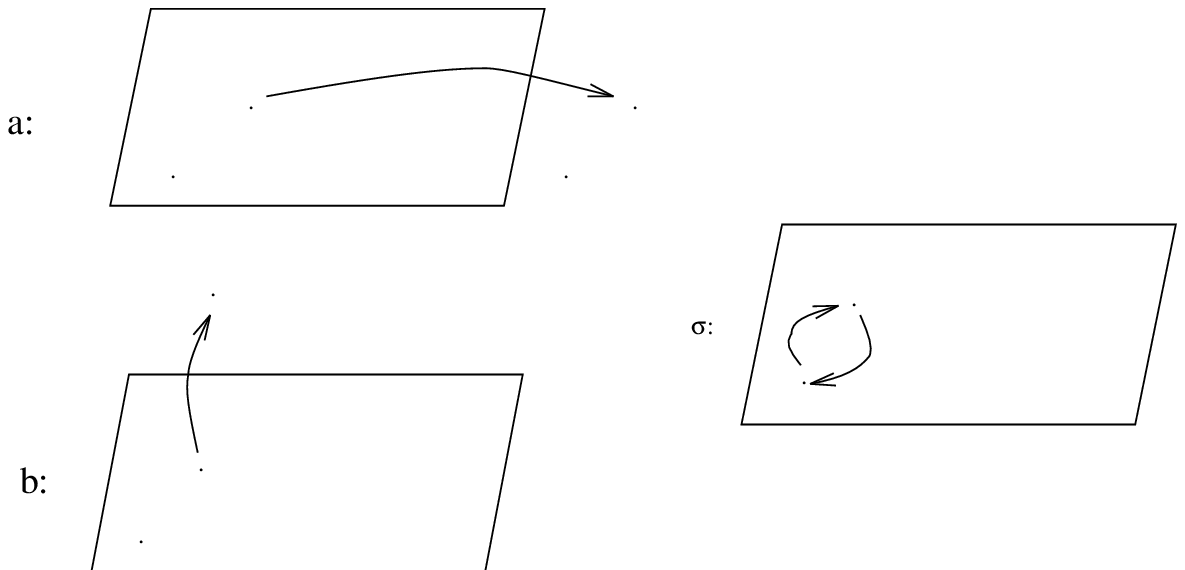}

These generators inject into ${\cal M}_{1,2}$ by doing twists in a 
neighbourhood of the above paths in such a way that in the non punctured torus they 
project to the identity.  That is one twists to bring the punctures along the paths 
indicated and then untwists so that once the punctures are forgotten, the mapping 
class group elements deform to the identity.  In \cite{B} a set of relations was derived 
for $B_{1,2}$.  Noting that the center corresponds to the translation of 
both punctures by an a or b cycle, then we have the following relations for 
$B_{1,2}/center$:  
\begin{equation}
\sigma^{-1}a\sigma^{-1}a=1,\; \; \sigma b\sigma b=1, \; \; aba^{-1}b^{-1}=\sigma^2. 
\end{equation}

We need also lifts of the generators of the mapping class group of the nonpunctured torus.
The best way to see these lifts is by the way they act on a set of generators of the fundamental group of the two punctured torus.  Here the generators chosen are the ones shown in the following diagrams:

\epsfxsize =\textwidth
\epsffile{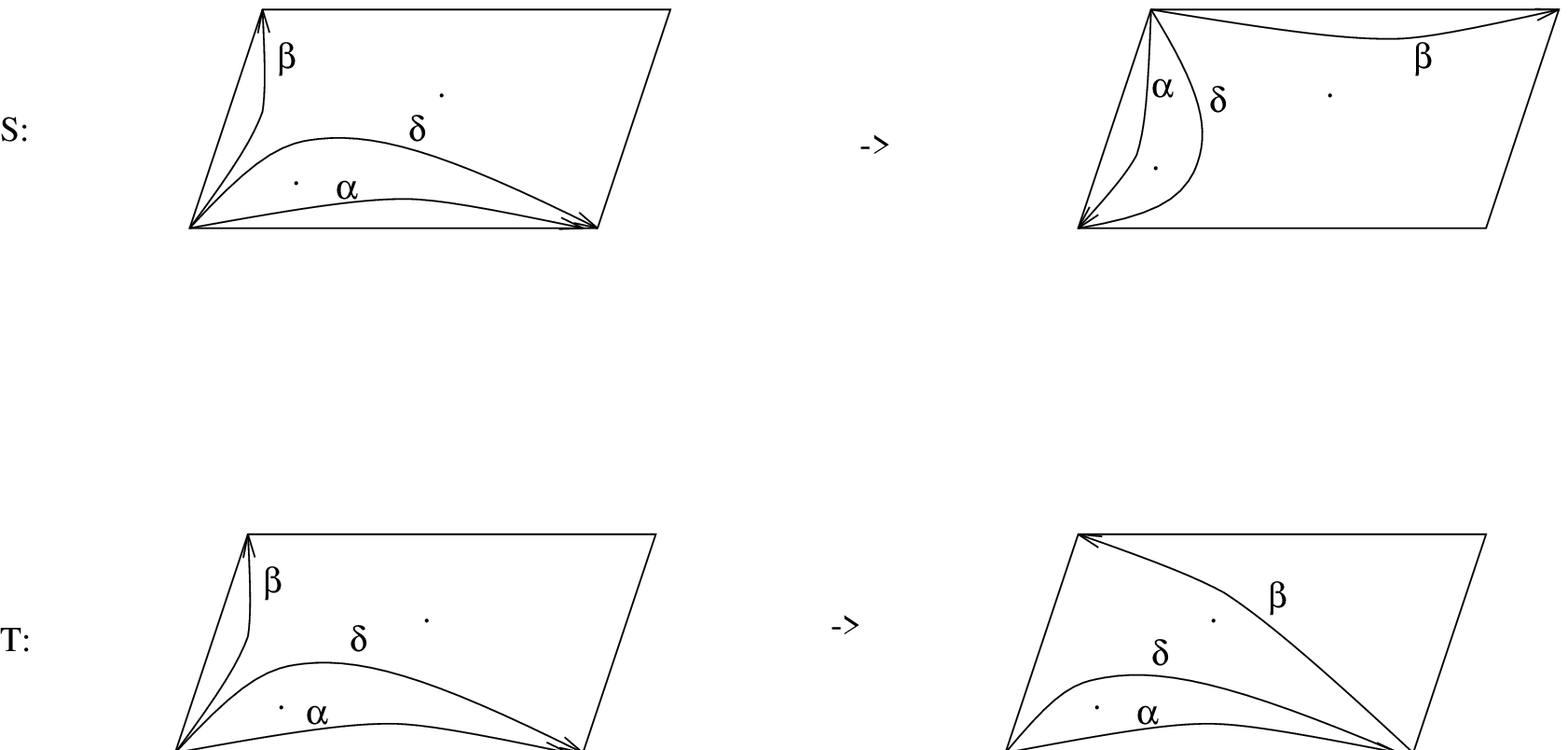}

The relations that these lifted generators $\tilde S$ and $\tilde T$ satisfy should project to the usual SL(2,Z) relations, that is $TSTSTS^{-1}=1$, $S^4=1$.  This means that the 
products $\tilde T\tilde S\tilde T\tilde S\tilde T\tilde S^{-1}$ and $\tilde S^4$ can be expressed in terms of $a,b,\sigma$.  
By considering the action of these products on the fundamental group it is easy to check that the relations become: 
\begin{equation}
\tilde T\tilde S\tilde T\tilde S\tilde T\tilde S^{-1}=1,\; \; \tilde S^4\sigma ^2=1.
\end{equation}
Here it is useful to remember that trivial elements of the mapping class group act by inner automorphisms on the fundamental group of the two punctured torus by moving the base point.  So two 
automorphisms of this fundamental group can be identified if they differ by an inner 
automorphism.  As an example we deduce the second of these relations:

\epsfxsize =\textwidth 
\epsffile {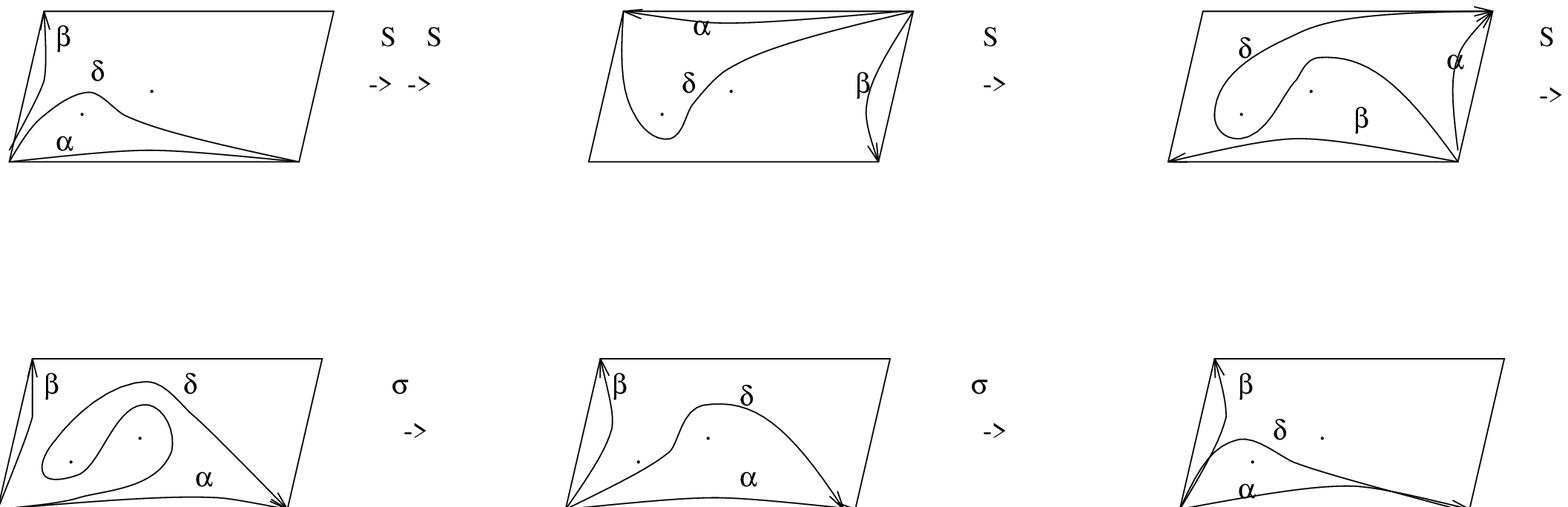}

Now to get the mixing relations we need only to consider the products $\tilde S^{-1}a\tilde S$, 
$\tilde S^{-1}b\tilde S$, $\tilde S^{-1}\sigma \tilde S$, $\tilde T^{-1}a\tilde T$, $\tilde T^{-1}b\tilde T$, $\tilde T^{-1}\sigma \tilde T$.  These 
project to 1, so they should be expressible in terms of $a,b,\sigma $.  To see  
why it is sufficient to consider these expressions consider an identity of the form word=1.  We can use the above relations to collect $\tilde S$ and $\tilde T$ on one side of the word. 
Upon projection only $\tilde S$ and $\tilde T$ survives, and the projected product is 1.  So we can use the lifted SL(2,Z) identities to express this product in terms of the braid group generators. 
Since there is not any further relation satisfied by the projected S and T, this is always possible.  So we are left with a braid generators relation and we know a complete set of 
such relations.  So the above products give the only new relations of our generators.  
It is not very difficult to show that the new relations we get are the following:
\begin{equation}
\tilde S^{-1}b\tilde S=a^{-1},\; \; \tilde S^{-1}a\tilde S=\sigma^2b, \; \; \tilde S^{-1}\sigma \tilde S=\sigma, 
\end{equation}
\begin{equation}
\tilde T^{-1}b\tilde T=ba, \; \; \tilde T^{-1}a\tilde T=a, \; \; \tilde T^{-1}\sigma \tilde T=\sigma .
\end{equation}

\section{Identities satisfied by the transformation matrices}
Now lets go back to the transformation matrices of the conformal blocks.
The way to construct identities of the transformation matrices is to consider 
contractible loops in the z-plane.  Here the blocks were described as densities 
on the universal cover of the torus with branch points at the points $m+n\tau $. 

First we cut the plane at the points $n\tau $ horisontally to the right 
and at the points $1+n\tau $ horizontally to the left.  In this way our blocks are 
uniquely defined on the cut plane.   

Note that we have the matrix b associated to the analytic 
continuation $z\rightarrow z-\tau $ and this is different from the matrix 
associated to the transformation $z+\tau \rightarrow z$ because by continiously 
moving z we cannot move the straight path from $z$ to $z-\tau $ to the straight 
path from $z+\tau $ to $z$ without crossing a branch point.  So let us denote the 
matrix associated to the second transformation by $\hat b$.  We have a 
transformation matrix that allows us to move around the origin and this is the 
$\sigma $ matrix.  In fact $\sigma ^2$ will give us the monodromy matrix for 
carrying $z+\tau $ clockwise, through the cut from 0 to $-\infty +0i$, back to $z+\tau $. So we have $\hat b=\sigma^2b\sigma^{-2}$.  
 
Consider now the following diagrams:

\epsfxsize =\textwidth
\epsffile{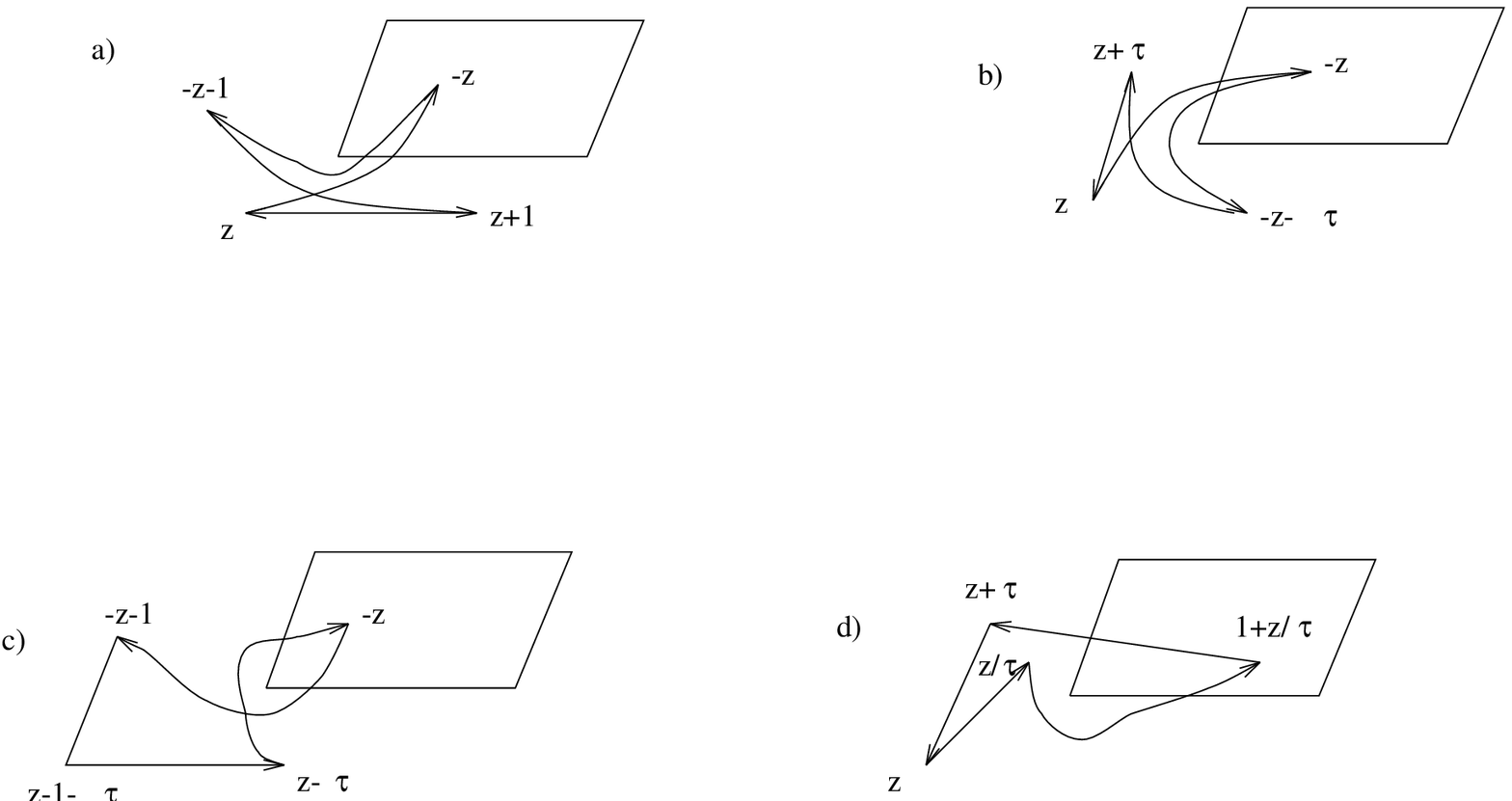}

The first three diagrams correspond to the following identities:
$\sigma ^{-1}a\sigma ^{-1}a=1$, $\sigma b\sigma b=1$, 
$\sigma ^2=aba^{-1}b^{-1}$.   
The fourth diagram corresponds to the identity $S^{-1}a^{-1}S\hat b=\sigma^{-2}$ 
Also, noticing that doing twice the transformation 
$S:(z,\tau )\rightarrow (z/\tau ,-1/\tau )$ corresponds to the transformation 
$\sigma :z\rightarrow -z$ we get the identity $\tilde q^{3/4}\sigma =S^{-2}$.
The factor $\tilde q^{3/4}$ appears because the blocks are really densities, 
and in the S matrix we have already taken this into account.  
The last two identities relate the S and T transformations to the braiding generators.  There is also one more relation.  This is the relation $TSTST=S$ satisfied by 
S and T transformations as can be seen again by analytic continuation arguments.    
These identities have been tested for levels k=1,2,3 for the matrices given 
in the previous section.  

Finally, it is not difficult to check that if we let $\tilde S=\tilde q^{3/8}S$ and 
$\tilde T=\tilde q^{-1/8}T$ the mapping class group relations are satisfied.  So we have constructed an infinite class of representations of ${\cal M}_{1,2}$ acting 
on the space of two point conformal blocks on the torus.  

{\bf Acknowledgements:}
I would like to thank prof. Looijenga for his insight into the structure of Mapping Class Groups.  I would like also to thank Dr. Vosegaard for useful discussions.

\end{document}